\documentstyle[twoside,fleqn,espcrc2,epsf]{article}

\newcommand{\Dslash}{$D$\kern-0.6em \hbox{/}}

\newcommand{\la}{\raise.16ex\hbox{$\langle$}}
\newcommand{\ra}{\raise.16ex\hbox{$\rangle$}}
\newcommand{\be}{\begin{equation}}
\newcommand{\ee}{\end{equation}}
\newcommand{\bea}{\begin{eqnarray}}
\newcommand{\eea}{\end{eqnarray}}

\title{The relevance of center vortices
\thanks{Talk presented by Ph. de Forcrand}}

\author{
C. Alexandrou\address{University of Cyprus, CY-1678 Nicosia, Cyprus and PSI,
 CH-5232 Villigen, Switzerland},
M. D'Elia\address{University of Pisa and INFN, I-56127 Pisa, Italy}
and Ph. de Forcrand\address{ETH, CH-8092 Z\"urich, Switzerland} }

\begin{document}

\begin{abstract}
We show remnants of chiral symmetry breaking in the center-projected theory.
We construct and study an unambiguous definition of center vortices.
\end{abstract}
\maketitle

\vspace*{-0.9cm}

\section{Non-perturbative effects in the center-projected theory}

The standard approach to identify center vortices proceeds through 
gauge fixing. 
In \cite{PRL}, we fixed an $SU(2)$ ensemble to
Direct Maximal Center (DMC) gauge, by iteratively maximizing
\be
Q(\{U_\mu\}) \equiv \sum_{x,\mu} \left( \mbox{Tr}~U_\mu(x) \right)^2 \; . 
\label{max}
\ee
After factorizing gauge-fixed links as
$U_\mu(x) = {\rm sign}(\mbox{Tr}~U_\mu(x)) \times U'_\mu(x)$
we studied the properties of the  ensemble $\{U'_\mu(x)\}$,
which by construction contains no center vortices. We showed that all
non-perturbative features had disappeared: confinement, chiral symmetry
breaking, and non-trivial topology. We have now looked at the center-projected
theory $\{\rm{sign}(\mbox{Tr}U_\mu(x))\}$, to see if it inherits the 
non-perturbative
properties of the original. As in \cite{sigma}, we observe
that the string tension is consistent with its $SU(2)$ value. 
We measure 
the quark condensate $\langle \bar{\psi} \psi \rangle(m_q)$ as a probe
of chiral symmetry breaking. Fig.1 shows that it clearly extrapolates 
linearly to a non-zero value as $m_q \rightarrow 0$. Furthermore, it diverges
as $1/m_q$ for very small quark masses, revealing the presence of a few
extremely small eigenvalues which may be caused by the non-trivial topological
content of the $SU(2)$ configuration. Note the similarity of Fig.1 with
the quenched condensate observed with domain-wall fermions \cite{DW}.
However, the associated quasi-zero modes appear to be strongly localized
and not chiral (i.e., $\bar{\psi} \gamma_5 \psi \sim 0$).

\begin{figure}[t]
\begin{center}
\vspace*{0.5cm}
\epsfxsize=6.5cm 
\epsfysize=4.5cm 
\epsffile[57 195 571 669]{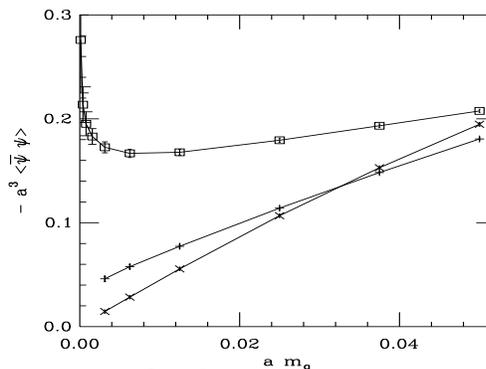}
\end{center}
\vspace{-2.5cm}
\caption{Quark condensate versus quark mass, in the
  original (+),
the center-vortex free (x), and the center-projected (squares) theories.}
\end{figure}

\vspace*{-1.5cm}

\section{Unambiguous center-vortex cores}

The local implementation of the DMC gauge-fixing  Eq.(\ref{max}) is
ambiguous, leading to 
many local maxima. The properties of $P$-vortices
obtained from different Gribov copies can be dramatically 
different \cite{Kovacs}. Here we define an unambiguous gauge condition.
Note first that DMC is equivalent to maximizing 
$\sum_{x,\mu} {{\rm Tr_{adj}}}~U_\mu(x)$
since 
${{\rm Tr_{adj}}}~U = 2 \left( \rm{Tr}~U \right)^2 - 1$.
The idea is thus to smooth the center-blind, adjoint component of the
gauge field as much as possible, then to read the center component off
the fundamental gauge field. Therefore, Maximal Center Gauge is just another
name for adjoint Landau gauge.

The problem of Gribov copies in the fundamental Landau gauge was solved
in \cite{VW} by  computing the covariant Laplacian
$\Delta_{xy} = 2d \delta_{xy} - \sum_{\pm \hat{\mu}} U_{\pm \hat{\mu}}(x)
\delta_{x \pm \hat{\mu},y}$
and its lowest-lying eigenvector $\vec{v}$. At each site,
$v(x)$ has 2 complex color components. The Laplacian gauge condition consists
of rotating $v(x)^\dagger$ along direction $(1,1)$ at all sites.
We  follow this construction for the adjoint representation. The covariant
Laplacian is now constructed from adjoint links
$U^{ab} = \frac{1}{2} \mbox{Tr}~[U \sigma^a U^\dagger \sigma^b],~a,b=1,2,3$.
It is a real symmetric matrix. The lowest-lying eigenvector $\vec{v}$ has
3 real components $v_i, \> i=1,2,3$
at each site $x$. One can apply a local gauge transformation $g(x)$ to rotate
it along some fixed direction, e.g., $\sigma_3$.
Note, however, that this does not specify the gauge completely: Abelian
rotations around this reference direction are still possible. Here
they are of the form $e^{i \theta(x) \sigma_3}$.
What we have achieved at this stage is a variation of Maximal Abelian Gauge
which is free of Gribov ambiguities. This Laplacian Abelian Gauge has been
proposed in \cite{AvdS}, which also shows that monopoles can not only be
identified through the DeGrand-Toussaint procedure in the Abelian projected
theory, but should be directly identifiable by the condition $|v(x)| = 0$
for smooth fields. Abelian monopole worldlines appear naturally as the locus
of ambiguities in the gauge-fixing procedure: the rotation to apply to 
$v(x)$ cannot be specified when $|v(x)| = 0$.

To fix to center gauge, we must go beyond Laplacian Abelian Gauge and specify
the Abelian rotation $e^{i \theta(x) \sigma_3}$. This is done most naturally
by considering the second-lowest eigenvector $\vec{v'}$ of the adjoint
covariant Laplacian, and requiring that the plane $(v(x),v'(x))$ be parallel
to, say, $(\sigma_3,\sigma_1)$ at every site $x$. This fixes the gauge
completely, except where $v(x)$ and $v'(x)$ are collinear. Collinearity 
occurs when $\frac{v_1}{v'_1} = \frac{v_2}{v'_2} = \frac{v_3}{v'_3}$,
i.e. 2 constraints must be satisfied. Thus, gauge-fixing ambiguities have
codimension 2: in 4$d$, they are 2$d$ surfaces. They
can be considered as the center-vortex cores for the following reasons:

First, note the analogy with fluid dynamics: in that context, a vortex refers
to a helical flow, with the vortex core at the center of the helix.
At the core, the centripetal acceleration vanishes, so that velocity and 
acceleration are collinear. Indeed, this collinearity condition has been
used to identify vortex cores in 3$d$ fluid flow \cite{fluid}.

Consider now the intersection of our 2$d$ center-vortex core with some plane
$(\mu,\nu)$ at  a point $x_0$. As one describes a small loop around $x_0$
in the plane $(\mu,\nu)$, the rotation $\theta(x)$ necessary to maintain
$(v(x),v'(x))$ parallel to $(\sigma_3,\sigma_1)$ varies by $2 \pi$,
reflecting the gauge singularity at $x_0$. But this is a rotation of the
{\em adjoint} field, so that the gauge rotation of the fundamental field
as one goes around the small loop will be 
$e^{-i \frac{1}{2} 2\pi \sigma_3} = - \bf{1}$.
A small Wilson loop around $x_0$ will have trace $-1$ in the fundamental
representation. This shows that center-vortex cores are aptly named, 
since they are indeed dual to $-1$ small Wilson loops. If the gauge field
is smooth, these $-1$ loops will also be identified by the usual procedure,
consisting of extracting $\rm{sign}(\mbox{Tr}~U_\mu(x))$ and computing the
$Z(2)$ plaquette. The so-called $P$-vortices constructed that way are indeed 
almost dual to the center-vortex cores, but not exactly. This is because
they are obtained by a somewhat arbitrary, non-linear recipe. 
In our construction, unlike in DMC, the center-vortex cores where gauge fixing
is ambiguous are the fundamental objects.

Our Laplacian Center Gauge solves the Gribov problem. 
It does not require an underlying lattice, but
 can be studied in the continuum like the original
Laplacian gauge \cite{PvB}. And it exhibits the center-vortex cores as an
intrinsic property of the gauge field, independently of the gauge condition
chosen: one could specify to rotate $v(x)$ and $v'(x)$ at every site along
arbitrary, $x$-dependent directions rather than $\sigma_3$ and $\sigma_1$.
The center-vortex cores will be unchanged.\\
Still, some arbitrariness remains in two respects: \\
$(i)$ Another discretization of the covariant Laplacian could be used, with
higher-derivative, irrelevant terms.
This will
affect the location and density of the center-vortex cores. \\
$(ii)$ Another choice of covariant eigenvectors $\vec{v},\vec{v'}$ could
be made. While it seems natural to choose for $\vec{v}$ the lowest-lying
eigenmode to maximize the smoothness of the gauge, the choice of $\vec{v'}$
appears less crucial. The third eigenvector of the Laplacian could be taken
as well, or even the second eigenvector of a Laplacian modified as per $(i)$.
Under a different choice of $\vec{v'}$, the center-vortex cores will change,
but not the monopoles, defined by $|v(x)|=0$. The 1$d$ monopole worldlines
remain embedded in the 2$d$ center-vortex cores. In fact, one may view the
center-vortex cores as the sheet spanned by the Dirac string of the monopoles
in the adjoint representation. 

We have applied Laplacian Center Gauge fixing and center projection to an
ensemble of $SU(2)$ configurations. As in \cite{PRL}, 
the string tension, the quark condensate and the 
topological charge all vanish upon removal of the $P$-vortices. The $Z(2)$ string
tension is consistent with its $SU(2)$ value, and the $Z(2)$ quark condensate
behaves as in Fig.1. The main difference is that the $P$-vortex density 
is higher than with DMC ($\sim 11\%$ vs $\sim 5.5\%$). The same effect was
observed for the monopole density in Laplacian Abelian Gauge \cite{AvdS}.

\begin{figure}[!t]
\begin{center}
\epsfxsize=6.2cm 
\epsfysize=4.5cm 
\epsfclipoff
\epsffile[0 150 565 545]{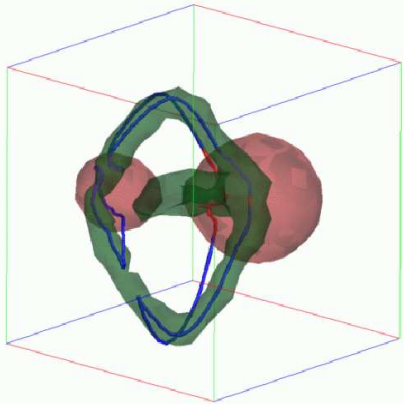}
\vspace{-0.7cm}
\caption{Time-slice of a cooled two-instanton (pink) configuration. The
blue loops are center-vortex cores. The green surface indicates high $Z(2)$
action density.}
\end{center}
\vspace{-1.8cm}
\end{figure}

\begin{figure}[!t]
\begin{center}
\epsfxsize=6.2cm 
\epsfysize=4.9cm 
\epsfclipoff
\epsffile[0 130 540 580]{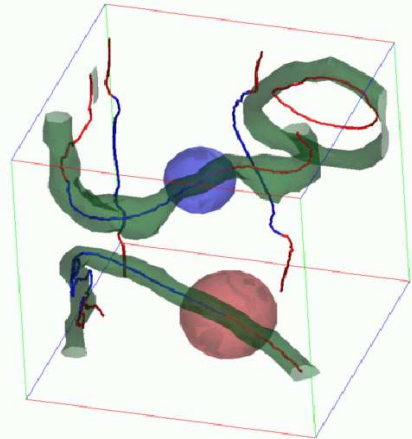}
\end{center}
\vspace{-1.5cm}
\caption{Time-slice of a cooled caloron configuration. The two monopoles are
pink and blue. The center-vortex cores (closed loops) pierce the monopoles
at their center (where $|v|=0$). The green surface indicates high $Z(2)$
action density.}
\vspace{-0.6cm}
\end{figure}


We have also applied our procedure to classical configurations. Fig.2 shows
a cooled two-instanton configuration. Note the double loop of vortex cores,
which shows interesting signs of self-intersection, as required in the
continuum \cite{Reinhardt}. The green area shows the $Z(2)$ action 
density, coming from $P$-vortices: while overall agreement is quite 
good between vortex cores and $P$-vortices, the latter are more sensitive
to UV fluctuations and show some spurious structure. 
Fig.3 shows a caloron,
i.e. a large instanton at finite temperature. As explained in \cite{caloron},
it decomposes into a pair of monopoles, identifiable by their action and 
topological charge densities. The center-vortex cores recognize these monopoles:
they pierce them and $|v|$ vanishes precisely at their center,
as shown by the change of color (red=$v,v'$ parallel; blue=anti-parallel).
Again the $P$-vortices give a slightly modified picture. 

Finally, our procedure readily generalizes to $SU(N)$: complete gauge-fixing
is achieved by rotating the first $(N^2-2)$ eigenvectors of the adjoint
Laplacian along some reference directions. Ambiguities arise whenever
these $(N^2-2)$ eigenvectors 
[each with $(N^2-1)$ real components] 
become
linearly dependent. This again defines codimension-2 center-vortex cores.


\end{document}